\documentclass[a4paper,fleqn,usenatbib]{mnras}

\usepackage{newtxtext,newtxmath}

\usepackage[T1]{fontenc}
\usepackage{ae,aecompl}

\usepackage[normalem]{ulem}

\usepackage{graphicx}	
\usepackage{amsmath}	
\usepackage{amssymb}	
\usepackage{lipsum}

\usepackage{array}
\usepackage{tikz}

\def\aj{AJ}%
\def\apj{ApJ}%
\def\apjl{ApJ}%
\def\apjs{ApJS}%
\def\ao{Appl.~Opt.}%
\def\aap{A\&A}%
\def\mnras{MNRAS}%
\def\nat{Nature}%
\title[Galaxy mergers do not lie on the FMR]
{Galaxy pairs in the Sloan Digital Sky Survey -- XIV. Galaxy mergers do not lie on the Fundamental Metallicity Relation}

\author[S.~Bustamante et al.]
{\parbox{17cm}
{Sebasti\'an~Bustamante$^{1,2}$, Sara~L.~Ellison$^{3}$, David~R.~Patton$^{4}$\\and Martin~Sparre$^{5,6}$}\vspace*{0.1cm}\\
$^{1}$Heidelberger Institut f\"{u}r Theoretische Studien,
  Schloss-Wolfsbrunnenweg 35, 69118 Heidelberg, Germany\\
$^{2}$Zentrum f\"ur Astronomie der Universit\"at Heidelberg, Astronomisches Recheninstitut, M\"{o}nchhofstr. 12-14, 69120 Heidelberg, Germany\\
$^{3}$Department of Physics and Astronomy, University of Victoria, Victoria, BC V8P 1A1, Canada\\
$^{4}$Department of Physics and Astronomy, Trent University, 1600 West Bank Drive, Peterborough, ON K9L 0G2, Canada\\
$^{5}$Institut f\"ur Physik und Astronomie, Universit\"at Potsdam, Karl-Liebknecht-Str.\,24/25, 14476 Golm, Germany\\
$^{6}$Leibniz-Institut f\"ur Astrophysik Potsdam (AIP), An der Sternwarte 16, 14482 Potsdam, Germany
 }

\pubyear{2020}

\begin{document}
\label{firstpage}
\pagerange{\pageref{firstpage}--\pageref{lastpage}}
\maketitle

\begin{abstract}
In recent observational studies, star-forming galaxies have been shown to follow a relation often dubbed the fundamental metallicity relation (FMR). This relation links the stellar mass of a galaxy with its star formation rate (SFR) and its gas-phase metallicity. Specifically, the FMR predicts that galaxies, at a given stellar mass, exhibit lower metallicities for higher SFRs. This trend is qualitatively consistent with observations of galaxy pairs, which have been robustly shown to experience increasing gas-phase metallicity dilution and enhanced star formation activity with decreasing projected separation. In this work, we show that, despite the qualitative consistency with FMR expectations, the observed O/H dilution in galaxy pairs of the Sloan Digital Sky Survey is stronger than what is predicted by the FMR. We conclude that the evolutionary phase of galaxies interacting with companions is not encoded in the FMR, and thus, mergers constitute a clearly defined population of outliers. We find that galaxies in pairs are consistent with the FMR only when their separation is larger than 110 kpc. Finally, we also quantify the local environment of the pairs using the number of galaxy neighbours within $2\, \mathrm{Mpc}$, $N_2$, and the projected separation to the second closest galaxy, $r_2$. We find that pairs are more sensitive to a second companion than to the local galaxy density, displaying less elevated SFRs with smaller values of $r_2$.
\end{abstract}

\begin{keywords}
galaxies: interactions, galaxies: star formation, galaxies: evolution, galaxies: abundances
\end{keywords}

\section{Introduction} \label{SecIntro}

The cycle of life and death of stars leads to a steady enrichment of heavy elements in the interstellar medium (ISM), and together with processes such as accretion of low-metallicity gas from reservoirs around galaxies and gas outflows induced by stellar and AGN feedback, yields a correlation between the stellar mass ($M_{\star}$) and the metallicity\footnote{Throughout this paper, the term \emph{metallicity} will be employed to refer to the gas-phase metallicity, unless otherwise stated.} of the star-forming gas \citep{Tremonti-2004}. However, the scatter of the relation is significant \citep[$\sim 0.1\ \mathrm{dex}$;][]{Tremonti-2004}, which means that galaxies, although building up their content of heavy elements gradually and steadily during most of their lifetime, are also evolving on diverse pathways through the stellar mass-metallicity plane.

If the role of stellar feedback is important in regulating the metal content of the ISM, and gas infall drives star formation activity, a secondary dependence on the SFR is expected in the mass-metallicity relation. This was first discovered by \citet{Ellison-2008}, where a weak dependence on the specific SFR (SSFR) was found for SDSS galaxies. Specifically, galaxies at a fixed stellar mass exhibit lower metallicities for higher SFRs. This relation is often dubbed the \textit{fundamental metallicity relation} (FMR). More recent studies have also confirmed the presence of the FMR, both in observations \citep{Mannucci-2010, Lara-Lopez-2010, Belli-2013, Stott-2013, Yabe-2015} and simulations \citep{Lagos-2016, De-Rossi-2017, Dave-2017, Torrey-2017, Bustamante-2018}. However, it is still debated whether the FMR is \emph{fundamental}, i.e. whether it is redshift independent and universal. Several observations at $z\sim 2.3$ have revealed differences from the FMR in the local Universe \citep{Salim-2015, Grasshorn-2016, Brown-2016, Sanders-2018} suggesting	 that the redshift-dependence of the FMR can be explained by an evolution of the metallicity of infalling gas and the mass-loading factor at fixed stellar mass. \citet{Ellison-2011} find that barred galaxies have both high SFRs and high metallicities, which is opposite to the FMR expectations. Taken together, the observations suggest that there are at least some galaxy populations that do not follow the FMR.

Interestingly, the predictions of the FMR are qualitatively in good agreement with observations of galaxy pairs, in which systematic offsets towards lower gas-phase metallicities and higher SFRs with respect to control samples of isolated galaxies have been previously reported \citep{Ellison-2008B, 2008MNRAS.386L..82M, Rupke-2010, Scudder-2012, Cortijo-Ferrero-2017, Thorp-2019}. This is also supported by numerical studies, in which metallicity dilution and SFR enhancement have been shown to take place in interacting galaxies as a result of merger-induced nuclear inflows \citep{Rupke-2010B, Torrey-2012, Hani-2018, Bustamante-2018}.

Nevertheless, a more careful statistical analysis of SDSS galaxies has revealed an overabundance of low-metallicity outliers in the FMR compared to a Gaussian distribution of residuals. \citet{Gronnow-2015} conclude that interacting galaxies might be at the origin of this population of outliers, as strong merger-induced nuclear inflows might cause substantial metallicity dilutions that are largely underestimated by the FMR. \citet{Bustamante-2018} (B18 henceforth) have used a set of cosmological zoom-in simulations to show that the FMR does not describe the evolutionary phases of merging galaxies, including post-merger stages. All of this further supports the idea that the FMR is not fundamental in general, and that mergers in particular may be outliers.

In this paper, we study the metallicity dilution and SFR enhancement of galaxy pairs and post-merger galaxies from the Sloan Digital Sky Survey Release 7 (SDSS DR7; \citealt{Abazajian-2009}). 
We use galaxy pairs compiled by \citet{Patton-2016} and post-merger galaxies compiled by \citet{Ellison-2013} and \citet{Ellison-2015}. We construct statistical control samples following the procedure proposed by \citet{Patton-2016}, i.e. matching simultaneously in stellar mass, redshift, local density and isolation. This allows us to quantify differences between pairs and post-mergers and their respective control samples over a wide range of mass ratios, projected separations and environments. Additionally, we fit a FMR to our whole galaxy sample in order to compare the predicted metallicity dilution and SFR enhancement to the observed values in galaxy pairs and post-mergers.

This paper is structured as follows. In Section~\ref{SecSample}, we present our sample selection and the matching algorithm for the control galaxies. In Section~\ref{SecDataAnalysis}, we fit a FMR to our galaxy sample and quantify the differences between the predicted and observed values of the metallicity dilution and the SFR enhancement for galaxy pairs and post-merger galaxies. Furthermore, we study the dependence of these processes on the stellar mass ratio and the local environment. Finally, we discuss our results in Section~\ref{SecDiscussion} and conclude in	 Section~\ref{SecSummary}. Throughout this work, we adopt a concordance cosmology with $\Omega_\Lambda = 0.7$, $\Omega_\mathrm{M} = 0.3$ and $H_0 = 70\ \mathrm{km\, s}^{-1}\mathrm{Mpc}^{-1}$.

\section{Sample selection} \label{SecSample}

\subsection{ Galaxy sample } \label{SubsecGalSam}

In the following, we describe the selection criteria applied to construct our general sample of galaxies, from which we also compile the samples of galaxy pairs, post-merger galaxies and controls. Our starting point is the spectroscopic pool of galaxies in the SDSS DR7. Following \citet{Patton-2013}, we select galaxies with reliable spectroscopic redshifts (redshift confidence of $\mathrm{zConf}>0.7$) and extinction-corrected $r$-band Petrosian apparent magnitudes in the range of $14.0\leq m_r \leq 17.77$. The redshift range is limited to $0.005 < z < 0.2$ in order to avoid spurious effects from the extremes of the redshift distribution. Each galaxy must have a reliable estimate of the total stellar mass from \citet{Mendel-2014}. 

Due to our interest in studying merger-induced effects on the SFR and the metallicity, we focus on those galaxies with reliable measurements of these quantities from the MPA catalogue. Additionally, we select only galaxies that are classified as star-forming using the criteria defined by \citet{Kauffmann-2003}.  Star formation rates were calculated by \citet{Brinchmann-2004}, based on H$\alpha$ fluxes with colour-based aperture corrections. We use global values of stellar mass and SFR throughout this work.

For the metallicities, we follow the selection methods and quality control cuts presented in \citet{Scudder-2012a}.  In brief, these include a S/N$>$5 requirement on the emission line fluxes of [OIII]$\lambda$5007, [OII]$\lambda$3727, [NII]$\lambda$6584, H$\alpha$ and H$\beta$ as well as cuts on the continuum error and raw flux errors.  The metallicities were computed using the adaptation of the \citet{Kewley-2002} method presented in \citet{Kewley-2008}. This method is calibrated against theoretical stellar population and photoionization models and uses an average of several $R_{23}$ calibrations for low metallicities and the ratio of [NII]/[OII] for high metallicities. An additional redshift cut of $z>0.02$ is performed in order to guarantee that all the emission lines used in the metallicity calibration are within the spectral range of the SDSS. We refer the reader to \citet{Scudder-2012a} for more details. 

For a given galaxy, the effects caused by the interaction with a close companion are in clear competition with environmental effects induced by the galaxy surroundings \citep{Cooper-2008,Ellison-2009,Peng-2012,Woo-2013}. Due to this reason, if one wants to understand the influence that the close companion has on the SFR and the metallicity, it is important to properly characterise both, the galaxy's closest companion and its host environment.  In order to do so, we characterise the closest potential companion of each galaxy by using the projected physical separation $r_p$, the rest-frame line-of-sight relative velocity $\Delta v$ and the stellar mass ratio (galaxy-to-companion stellar mass ratio) $\mu$. For the environment, we follow the approach presented in \citet{Patton-2016}, i.e. we use two different metrics to characterise a galaxy's environment: first, the total number of detected spectroscopic companions within a projected separation of $2\,   \mathrm{Mpc}$, $N_2$. The second metric is the projected separation to the galaxy's second closest companion, $r_2$. All the relevant galaxy companions must satisfy that their rest-frame line-of-sight velocity $\Delta v$ is within $1000\ \mathrm{km}\ \mathrm{s}^{-1}$ relative to the galaxy in question and that their corresponding stellar mass ratio is $\mu>0.1$, i.e. the total stellar mass of a galaxy is at least $10\%$ of the stellar mass of the primary galaxy. The combination of these two metrics allows us to distinguish between different types of environments. We roughly classify them into four categories: compact groups ($N_2<15$ and $r_2<150\,   \mathrm{kpc}$), galaxy clusters ($N_2\geq 15$ and $r_2<150\,   \mathrm{kpc}$), loose groups ($N_2\geq 15$ and $r_2\geq 150\,   \mathrm{kpc}$) and low density fields ($N_2< 15$ and $r_2\geq150\,   \mathrm{kpc}$). We refer the reader to \citet{Patton-2016} for more technical details and a through discussion on environment classification. After applying all the previous criteria, we end up with 68942 galaxies in our general sample.

\subsection{ Pair sample } \label{SubsecPairSam}

Although each galaxy in our general sample has a detected closest companion, if one wants to build a sample of galaxy pairs, it would be ideal that each companion has an appreciable physical influence on the galaxy in question, i.e. the pairs are not a projection artifact. In order to minimize the presence of projection artifacts, we impose the following conditions: the stellar mass of the companion is at least $10\%$ of the stellar mass of the main galaxy, i.e. $\mu>0.1$; less massive companions are likely to exert effects that are too small to be observed. Note that this criterion also allows companions that are more massive, i.e. $\mu>1$. The second condition restricts the relative rest-frame line-of-sight velocity $\Delta v$ to be within $300\, \mathrm{km}\, \mathrm{s}^{-1}$ of the main galaxy, thereby excluding galaxies with unrelated background and foreground companions \citep{Patton-2000}. This velocity cut is fairly standard and has been used in previous studies of galaxy pairs \citep{Ellison-2008B, Scudder-2012, Patton-2016} and a cut of this magnitude is supported by cosmological simulations \citep{Ventou-2019,Pfister-2020}. After applying all the previous criteria, our sample of pairs comprises $6031$ galaxies.

\subsection{ Post-merger sample } \label{SubsecPMSam}

In contrast to the selection of galaxy pairs, which uses an objective set of criteria based on angular and velocity separation, the identification of post mergers relies on visible features of disturbance.  Traditionally, this has been done using visual selection \citep[e.g.][]{Darg-2010, Simmons-2017} or quantitative morphological metrics \citep[e.g][]{Conselice-2003, Lotz-2004, Pawlik-2016}, although neural network applications are also now becoming popular \citep[e.g.][]{Bottrell-CNN,Pearson-2019,Walmsley-2019}. 

In the work presented here, we make use of the existing sample of post-mergers compiled by \citet{Ellison-2013} from the post-merger catalogue of \citet{Darg-2010}. This catalogue is based on a visual classification performed by the Galaxy Zoo project on SDSS galaxies and takes into account the presence of tidal features, strong irregularities, and signs of recent interaction. \citet{Ellison-2013} applied additional criteria to filter out irregular galaxies without signs of a recent interaction and galaxies that are undergoing a merger, but have not fully coalesced, e.g. a companion can still be detected. Additionally, the parent post-merger sample also contains the merger classifications described in \citet{Ellison-2015}, as well as some more recent visual classifications, resulting in $328$ post-merger galaxies from these various sources. The final post-merger sample is obtained by selecting those galaxies in the parent sample that are star-forming according to the \citet{Kauffmann-2003} criterion. This criterion yields $97$ post-merger galaxies, which are the galaxies used throughout this work.

The observability time-scale of merger features depends on several factors, such as the mass ratio of the merger and its gas fraction \citep{Lotz-2010a,Lotz-2010b}.  Identification of morphological asymmetries is also sensitive to the imaging depth of the survey \citep{Bottrell-S82}.  For example \citet{Ellison-2019} found that the frequency of disturbed galaxies was a factor of two higher in deep imaging from the Canada France Imaging Survey (CFIS) compared with SDSS.  We therefore expect that, qualitatively, the fairly shallow imaging of SDSS likely captures relatively major, gas-rich mergers quite recently after their final coalescence.  Based on the low metallicities observed in the original sample, \citet{Ellison-2013} estimate that coalescence has likely occurred within the last few hundred Myr.  \citet{Hani-2020} have recently identified post-mergers in the IllustrisTNG simulation and show that star formation rates return to normal within $\sim$ 500 Myr after coalescence.  The elevated SFRs of our post merger sample (reported already by \citealt{Ellison-2013}) are therefore again consistent with a relatively young post-merger age.

\subsection{ Control samples }\label{SubsecConSam}

Star forming galaxies follow two relationships that have been extensively studied and are now well-established; namely, the mass-metallicity relation \citep{Tremonti-2004, Ellison-2008} and the SFR-stellar mass main sequence \citep{Salim-2007, Noeske-2007, Speagle-2014}. These relations show that galaxies build up their heavy element content and stellar component in a steady fashion \citep[e.g. as modelled in various analytical galaxy evolution models;][]{Lilly-2013, Mitra-2015} as they grow in mass. Nevertheless, considering that galaxy mergers are not steady processes, an interesting question that arises is to what extent do mergers influence the evolution of the metallicity and star formation rate of galaxies. For individual system observations, such as our SDSS galaxy sample, only a single point in time can be probed, thereby making it difficult to study the evolution of any property. Alternatively, a statistical approach to the problem can be adopted. For instance, we construct control samples of isolated galaxies as a comparison point to quantify deviations in galaxy pairs and post-merger galaxies.

The SF main sequence and the mass-metallicity relation exhibit a redshift dependence, meaning that galaxies at different epochs behave differently \citep{Savaglio-2005, Miaolino-2008, Lara-Lopez-2010B, Whitaker-2012, Zahid-2014, Tasca-2015}. From the technical side, some aspects of the surveys such as sample depth, survey volume, spatial resolution and covering fraction also depend on redshift. This indicates that galaxies of the control sample must be simultaneously matched in stellar mass and redshift in order to reduce selection effects.

Although the influence of a close companion on a galaxy's properties is dominant at small projected separations, at larger separations the influence exerted by other neighbouring galaxies might be comparable or even larger \citep{Park-2007, Ellison-2010, Moreno-2013, Sabater-2013, Patton-2016}, constituting thus another source of selection effects. This is why the controls must be also matched in environment.

In this paper, we used the procedure presented by \citet{Patton-2016} to construct the control samples for pairs and post-mergers, in which a simultaneous matching in stellar mass, redshift and environment is performed. In order to match in environment, \citet{Patton-2016} employ the local galaxy density $N_2$ and an isolation criterion. In the case of galaxy pairs, isolation is guaranteed by requiring that the projected separation to the closest galaxy ($r_p$) of each of the controls is similar to the projected separation to the second closest companion ($r_2$) of the main member of the pair in question. In the case of post-mergers, $r_p$ is instead matched directly to the separation to the closest galaxy as, by definition, these systems have already coalesced.

The tolerances for each of the quantities that is being matched are taken from \citet{Patton-2016}. For the stellar mass we use $0.1\ \mathrm{dex}$, which is consistent with the statistical uncertainties from measurements  \citep{Mendel-2014}. For the redshift, the controls are required to agree to within $0.01$ with respect to the reference sample. Finally, for matching controls in environment, both quantities, $N_2$ and $r_p$, are required to be within $10\%$ of the value of the reference sample -- $r_2$ for galaxy pairs and $r_p$ for post-mergers when matching in isolation. This procedure gives at least $10$ control galaxies for $62\%$ of the galaxy pairs and $45\%$ of the post-merger galaxies. We increase iteratively all the tolerances by a factor of $1.5$ until we match at least 10 control galaxies for each of the remainders of the samples. Note that we are not excluding the case in which a galaxy might qualify as a control for multiple pairs and/or post-mergers.

\begin{figure*}
\centering
\includegraphics[width=0.70\textheight]{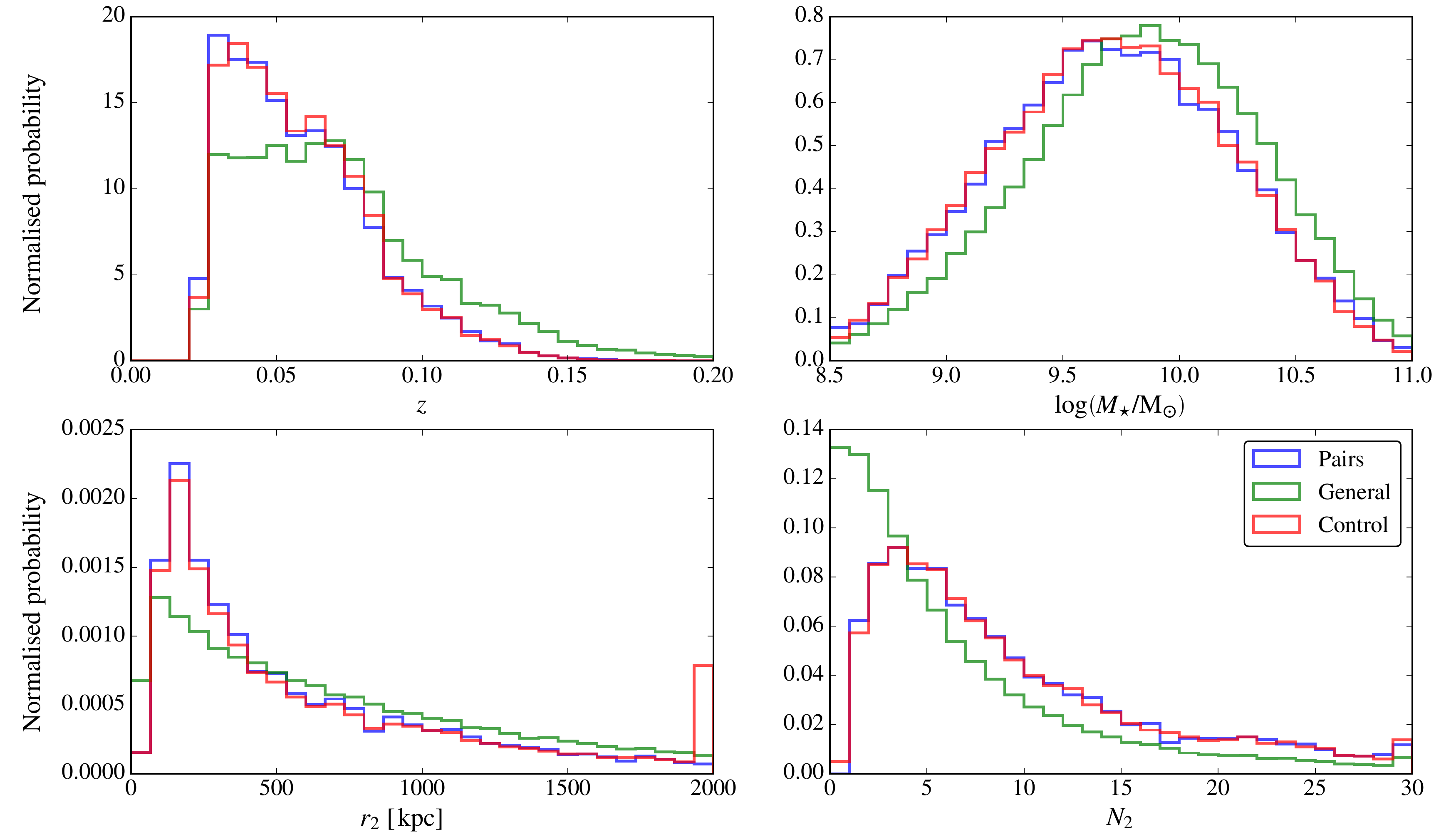}
  \caption{Normalised histograms of redshift (\emph{top left}), stellar mass (\emph{top right}), $r_2$ (\emph{bottom left}) and $N_2$ (\emph{bottom right}) for the galaxy pairs (blue). We also show the histograms for the general sample (green) and the controls (red). The distributions of the matching properties for galaxy pairs and their weighted controls are very well in agreement, with the only exception occurring at the high-end of the distribution of $r_2$. The reason for this is explained in the text (Section~\ref{SubsecConSam}). In the \emph{bottom left} panel, the histograms for the general sample and the controls denote $r_p$ (rather than $r_2$), since this is the quantity that is matched to the paired galaxy $r_2$ values.}
  \label{fig:ControlHistograms}
\end{figure*}

In order to select the best 10 matches for each galaxy, we adopt the weighting scheme used by \citet{Patton-2016}. For example, for the redshift of the $i$-th control galaxy, we define the corresponding weight as:
\begin{equation}
\label{eqn:Weights}
w_{z_i} = 1 - \frac{|z - z_i|}{z_{\mathrm{tol}}},
\end{equation}
where $z$ corresponds to the redshift of the galaxy for which the control is being matched. Analogous weights are defined for stellar mass, $N_2$ and $r_p$. The global statistical weight is then given by:
\begin{equation}
\label{eqn:TotalWeights}
w_{i} = w_{z_i}w_{M_i}w_{N_{2i}} w_{r_{2i}}.
\end{equation}

We select the 10 galaxies with the highest weights as the control set of a given pair or post-merger. In Figure~\ref{fig:ControlHistograms}, we show the histograms for the full sample of galaxies, galaxy pairs and their matched controls. In general, there is a very good agreement between the pairs and their controls, with only a deviation at the high-end of the distribution of $r_2$. The disagreement comes from the fact that $r_2$ and $r_p$ are capped at $2\,   \mathrm{Mpc}$ due to limited survey boundaries, and the metric $N_2$ that takes into account only companions within $2\,   \mathrm{Mpc}$ of the galaxy in question \citep{Patton-2016}. In order to circumvent potential selection issues coming from this, we discard galaxy pairs for which $r_2>1800\,   \mathrm{kpc}$. This only reduces the sample size by $7\%$. For the post-mergers, we also find a very good agreement with their controls, and the sample size is reduced by $15\%$ when applying the cut-off in $r_p$. 

In consideration of the above, the statistical mean of any quantity of interest $x$ for the control sample of a given galaxy can be computed as:
\begin{equation}
\label{eqn:WeightingQuantities}
\bar{x} = \frac{ \sum_{i=1}^{N_c} \omega_i x_i }{ \sum_{i=1}^{N_c} \omega_i },
\end{equation}
where $x_i$ is the measured quantity in the $i$-th control galaxy and $N_c$ the size of the control sample.

Once the control samples have been built, we can proceed to compute the relative differences of metallicity and SFR in galaxy pairs and post-mergers. None the less, there is a last issue that has to be accounted for before proceeding, i.e. spectroscopic incompleteness due to fibre collisions. As mentioned before, this observational artefact arises as a consequence of the finite size of optical fibres used in the survey, which creates a selection effect that disfavours the detection of companions within $55$ arcsec of any given galaxy. This selection effect impacts a different range of projected separations at any given redshift. In order to correct for this, we follow the approach introduced by \citet{Patton-2013} and described in detail by \citet{Patton-2016}, in which an additional statistical weight of $\omega_\theta = 3.08$ is assigned to galaxy pairs with projected separation smaller than $55$ arcsec; otherwise, $\omega_\theta = 1$. Another method commonly adopted to correct for this is a random culling of $67.5\%$ of the pairs with separations $>55\ \mathrm{arcsec}$ \citep[see for example ][]{Ellison-2008B}; however, this drastically reduces the sample size. 

\begin{figure*}
\centering
\includegraphics[width=0.72\textheight]{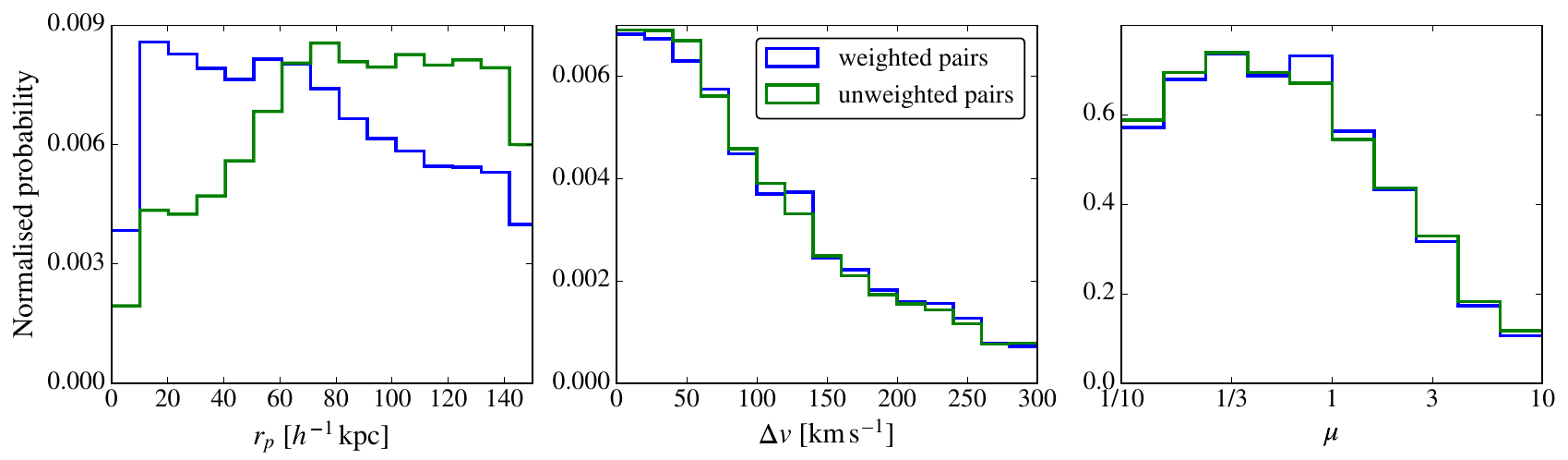}
  \caption{Normalised histograms of projected pair separation $r_p$ (\emph{left}), relative line-of-sight velocity $\Delta v$ (\emph{centre}) and stellar mass ratio $\mu$ (\emph{right}). Weighted (blue) and unweighted (green) histograms are shown. The use of statistical weights compensates for the deficit of galaxy pairs with projected separations $r_p<80\,   \mathrm{kpc}$. }
  \label{fig:PairHistograms}
\end{figure*}


In Figure~\ref{fig:PairHistograms}, we show the weighted and unweighted histograms for the properties of galaxy pairs, i.e. $r_p$, $\Delta v$ and $\mu$. Using the \citet{Patton-2013} fibre weighting scheme, the mean offset of any quantity $x$ for a given set of $N_g$ galaxies with respect to their controls can be computed as:
\begin{equation}
\label{eqn:FractionalChange}
\Delta x = \frac{ \sum_{j=1}^{N_g} \omega_{\theta_j} (x_j - \bar{x}_j) }{ \sum_{j=1}^{N_g} \omega_{\theta_j} },
\end{equation}
where $\bar{x}_j$ is the statistical mean of the controls according to equation~\ref{eqn:WeightingQuantities}.

\begin{figure}
\centering
\includegraphics[trim={0.1cm 2.5cm 2cm 3.cm},clip,width=0.34\textheight]{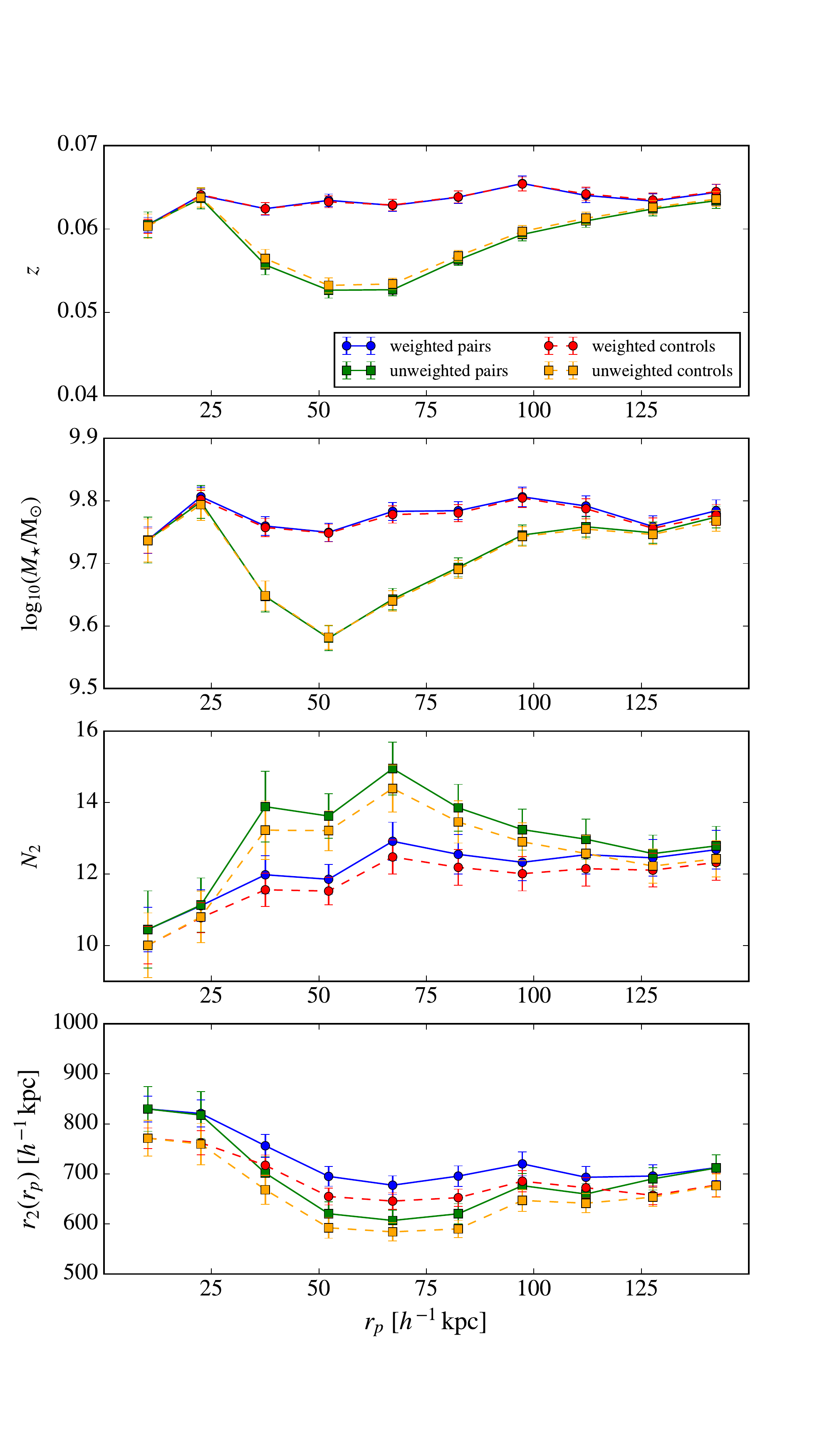}
  \caption{Mean redshift, stellar mass, $N_2$ and $r_2$ as a function of $r_p$ for galaxy pairs   (solid lines) and their controls (dashed lines). Error bars denote the standard error in the mean. Note that in the bottom panel, we show the distribution of $r_p$ for the controls and $r_2$ for the pairs. When comparing the weighted (circles) and unweighted (squares) distributions, the former exhibit more homogeneous trends in the range $25\,   \mathrm{kpc}<r_p<125\,   \mathrm{kpc}$. This shows that incompleteness due to fibre collisions is properly compensated.}
  \label{fig:RadialDependence}
\end{figure}

In Figure~\ref{fig:RadialDependence} we compute weighted and unweighted distributions of redshift, stellar mass, $N_2$ and $r_2$ as a function of $r_p$ for galaxy pairs and their controls. Our fibre weighting scheme is designed to remove the redshift bias of the pairs sample that would otherwise be present for pairs of intermediate separations, as described by \citet{Patton-2016}.  In particular, the minimum fibre angular separation of 55 arcsec and our sample redshift range of $0.02 < z < 0.2$ causes a redshift bias for pairs with $22.3 < r_p < 181.5$ kpc.
Within this range of $r_p$, fibre collisions will preferentially exclude the higher redshift pairs, biasing the sample to lower redshift. This bias can be seen clearly in the upper panel of Figure~\ref{fig:RadialDependence}. Weighted distributions exhibit noticeably more homogeneous trends in the range $25\,   \mathrm{kpc}<r_p<150\,   \mathrm{kpc}$, showing thus that fibre weights compensate for the deficit of close angular pairs. While small trends remain in the weighted distributions (e.g. a small increase in $N_2$ as $r_p$ increases), these trends are also present in the matched control sample, minimizing the influence of environment on our comparison of paired galaxies and their controls.

In summary, our full sample consists of 68942 galaxies, including 6031 paired galaxies and 97 post-mergers. 
We have also identified 10 control galaxies from the full sample for every paired galaxy and post-merger.

\section{Data Analysis}\label{SecDataAnalysis}

\begin{figure*}
\centering
\includegraphics[width=0.6\textheight]{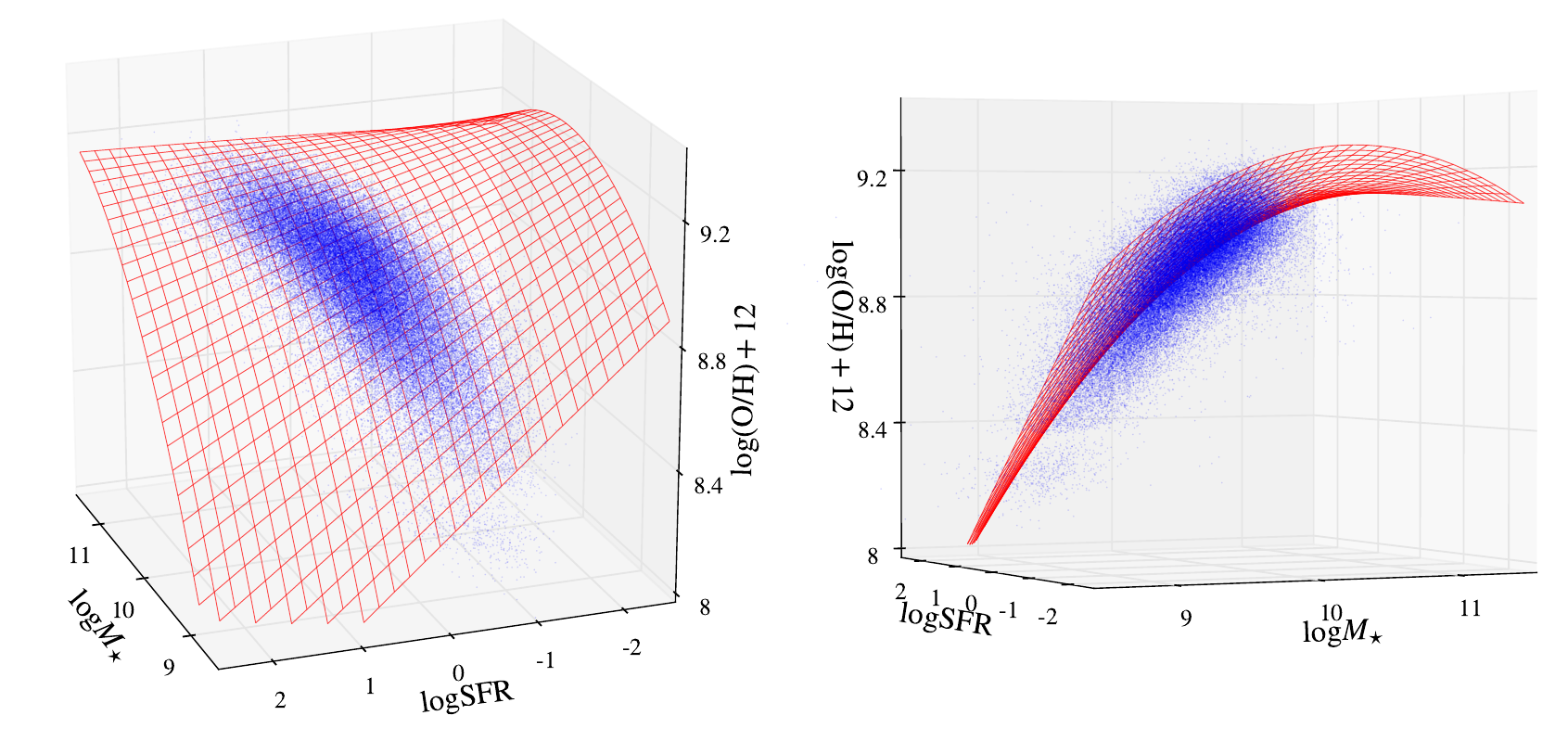}
  \caption{ Two projections of the distribution of SDSS galaxies in the three-dimensional space of stellar mass -- SFR -- metallicity. The red wireframe corresponds to the second-order polynomial fitted to the data and represents the FMR.}
  \label{fig:FMR3D}
\end{figure*}

\begin{figure}
\centering
\includegraphics[width=0.32\textheight]{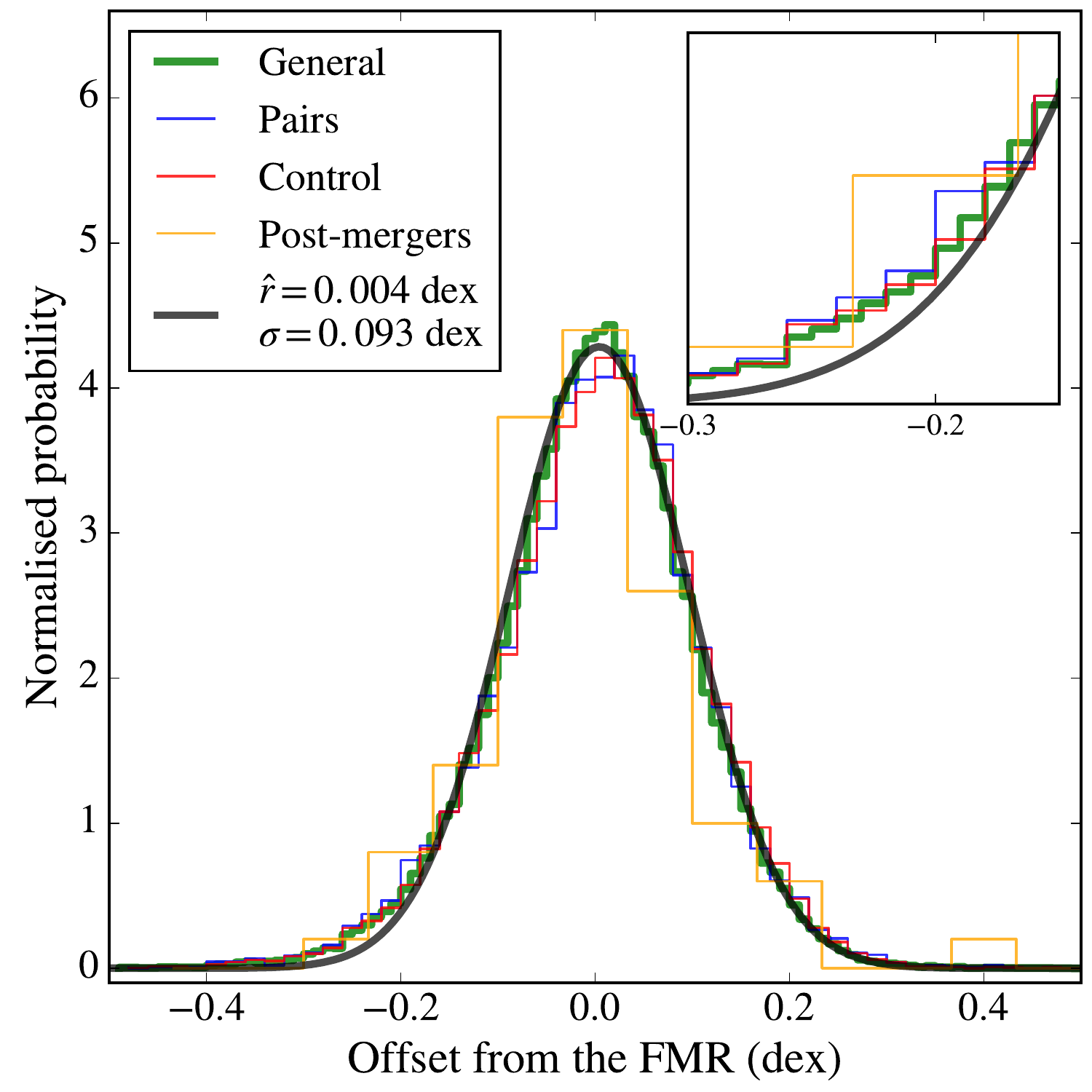}
  \caption{ Distributions of residuals of measured metallicity with respect to the fitted FMR for the general sample (green), galaxy pairs (blue), controls (red) and post-mergers (orange). The black line is a Gaussian fit to the residuals of the general sample with a standard deviation $\sigma = 0.093$. A clear overabundance of outliers in all samples occurs at the low-end of the distributions, approximately between $-0.4\ \mathrm{dex}$ and $-0.2\ \mathrm{dex}$, indicating lower metallicities than expected from the FMR.}
  \label{fig:Residuals}
\end{figure}

\subsection{ Fundamental metallicity relation } \label{SubsecFMR}

Our main goal is to investigate how galaxy pairs and post-mergers are positioned with respect to the FMR.  In order to characterise the FMR, we take the general sample of SDSS galaxies, in which galaxy pairs, post-mergers and their respective controls are included, and fit a second-order polynomial in $M_\star$ and SFR to the metallicity values. The polynomial represents the FMR. There are two different approaches to fit the data: the first one consists of binning the galaxies in $M_\star$ and SFR and then computing the metallicity median value in each bin; those values are subsequently used for fitting the FMR \citep{Mannucci-2010}. This approach reduces sampling issues as every bin is given the same statistical weight. The second approach is to fit the FMR directly to the \emph{raw} data \citep{Gronnow-2015}. We adopt the latter, but note that our results are qualitatively unchanged if we use the former approach. The fit yields:
\begin{eqnarray}
\nonumber
12 + \log( \mathrm{O}/\mathrm{H})_{\mathrm{FMR}}&=&\ 8.96 + 0.28 m - 0.12 s - 0.12 m^2 \\
\label{eqn:FMR}
&& \ + 0.085 ms - 0.001 s^2,
\end{eqnarray}
where $m \equiv \log(M_{\star}/\mathrm M_{\odot}) - 10$ and $s \equiv \log(\mathrm{SFR}/\mathrm M_{\odot}\mathrm{yr}^{-1})$. In Figure~\ref{fig:FMR3D} we show two projections, face-on and edge-on, of the distribution of SDSS galaxies along with the fitted FMR. Note that for stellar masses  $M_\star \lesssim 10^{10}\ \mathrm M_{\odot}$, the relation predicts that for lower SFRs, higher metallicity values are encountered at a fixed stellar mass. Although not visible in the projection shown in Figure~\ref{fig:FMR3D}, we note that beyond about $10^{10.5}\ \mathrm M_{\odot}$ the secondary dependence of O/H on SFR disappears and the form of the FMR changes such that it is no longer well described by equation~\ref{eqn:FMR}.  Our analysis is not significantly affected by this issue for two reasons.  First, the fraction of mergers in the affected mass range is small.  Second, our differential comparison between mergers and controls, where mass is one of the matched parameters, ensures that any bias will affect both samples equally.

With the aim of testing the goodness of fit of our model, represented by equation~\ref{eqn:FMR}, we compute in Figure~\ref{fig:Residuals} the distribution of residuals with respect to the FMR, defined as $r \equiv \log(\mathrm{O/H})_{\mathrm{data}} - \log(\mathrm{O/H})_{\mathrm{FMR}}$, for the general galaxy sample, the pairs, the post-mergers and the controls. For the general sample, we use $100$ bins uniformly distributed in $\log$ SFR and $\log M_*$ with a width of $0.05\ \mathrm{dex}$ to compute the distribution and then, using least squares, fit a Gaussian distribution function to it. This yields a mean value $\hat{r} = 0.004\ \mathrm{dex}$ and a standard deviation $\sigma = 0.093\ \mathrm{dex}$. Comparing with previous observational studies such as \citet{Mannucci-2010} and \citet{Gronnow-2015}, where standard deviations $\sigma = 0.053\ \mathrm{dex}$ and $\sigma = 0.048\ \mathrm{dex}$ are respectively computed, we find that our standard deviation is somewhat larger. We note that this does not necessarily imply a disagreement in the results regarding the FMR as there are several factors that can produce larger deviations in our data. For example, we are using the metallicity calibration of \citet{Kewley-2002}, whereas \citet{Mannucci-2010} and \citet{Gronnow-2015} use the calibration of \citet{Miaolino-2008} and \citet{Marino-2013}, respectively. As shown by \citet{Kewley-2008}, different calibrations can lead to different values of metallicity, with discrepancies such as the one we encounter within the range of what is expected. Furthermore, different parameter choices in the construction of the parent sample, such as S/N cuts and AGN culling, can also contribute to the different deviations. For the sake of consistency, we also compute the stellar mass - metallicity relation for the general sample. We find, for the distribution of residuals, a standard deviation of $\sigma = 0.16\ \mathrm{dex}$, which is larger than the value for our fitted FMR. This confirms that a secondary dependence of the metallicity on the SFR is also present in our data.

Finally, it is worth noting that all the distributions of residuals in Figure~\ref{fig:Residuals} exhibit a clear overabundance of outliers with respect to the Gaussian fit between $-0.4\ \mathrm{dex}$ and $-0.2\ \mathrm{dex}$. This corresponds to a small population of galaxies that are more metal poor than expected, based on the basic FMR. This feature has also been observed in previous works \citep[see e.g.][]{Mannucci-2010, Gronnow-2015} and has been proposed to be caused by the presence of interacting systems in the samples. However, in our sample, low metallicity outliers are seen in both the mergers and control samples.  This could either be caused by contamination of the control sample by unidentified mergers (since both spectroscopic selection of pairs in SDSS and of post mergers from visual identification are incomplete), or be due to secular processes contributing to FMR outliers. Further insight can therefore be obtained by a comparison directly between the mergers and their controls, an experiment we conduct in the next section.

\subsection{ Metallicity dilution and SF enhancement } \label{SubsecMetDilSFEnh}

\begin{figure}
\centering
\begin{minipage}{.05\linewidth}
\includegraphics[trim={0cm 0cm 34.5cm 0cm },clip,width=\linewidth]{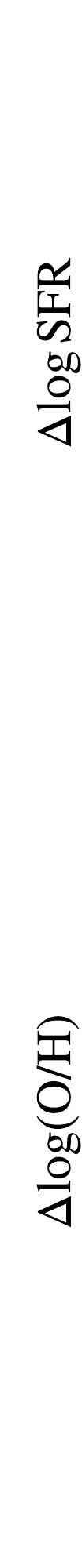}
\end{minipage}
\begin{minipage}{.93\linewidth}
\includegraphics[trim={18cm 0cm 0cm 0cm },clip,width=\linewidth]{figures/DilutionPairs.pdf}
\end{minipage}
  \caption{Offsets of SFR (\emph{top panel}) and metallicity (\emph{bottom panel}) for all $6031$ galaxy pairs as a function of projected separation. Data points show medians and error bars denote standard error of the median. Blue solid lines correspond to differences between controls and mergers for SFR or metallicity.  Green dashed lines show offsets of the pairs from the FMR (\emph{see text in subsection \ref{SecFMRTest} for details}) and red dotted lines show offsets of the controls from the FMR. At low projected separations, galaxy pairs deviate from the FMR-predicted values, exhibiting thus weaker SFR enhancement and stronger metallicity dilution, implying that the behaviour of close pairs is not well-described by the FMR.}
  \label{fig:DilutionPairs}
\end{figure}

We begin by computing the difference in SFRs and metallicities between the control and merger (pairs and post-mergers) samples.  We remind the reader that the mergers are matched to the controls in several quantities, such that the difference between them should capture the effect of the interaction.  In Figure~\ref{fig:DilutionPairs}, we show both differences in the SFR (top panel) and metallicity (bottom panel) between the mergers and control (blue line in both panels).  These results are independent of any characterisation of the FMR, and are analogous to previous studies that have quantified differences in SFR and metallicity in mergers and control samples.

In Fig. ~\ref{fig:DilutionPairs} we see a dependence of SFR enhancement and metallicity dilution on projected separation, with close pairs exhibiting larger offsets. For the sake of continuity, we show post-merger galaxies as an additional bin below $0$, as these systems have already coalesced. Our trends are consistent with those previously reported by \citet{Ellison-2013} and \citet{Scudder-2012}, who show that metallicity dilution becomes increasingly significant at separations $r_p \lesssim 60\, \mathrm{kpc}$ and that the trends continue to the post-merger sample. Our offsets at small separations are, however, somewhat weaker, i.e. $0.13\ \mathrm{dex}$ compared to $0.25\ \mathrm{dex}$ for SFR, and $0.035\ \mathrm{dex}$ compared to $0.048\ \mathrm{dex}$ for metallicity. The discrepancy in SFR might have its origin in the fact that we use total SFR quantities while \citet{Scudder-2012} use fibre SFR values. SFR offsets have been shown to be stronger in galaxy centres \citep[see e.g.][]{Patton-2011, Patton-2013, Ellison-2013, Thorp-2019}, and thus, the use of total quantities naturally leads to weaker values. Additionally, we note that the dependence of SFR enhancement and metallicity dilution on the pair mass ratio \citep[see subsection~\ref{SubsecMassRat} and][]{Scudder-2012} might also contribute to the discrepancy, as galaxy pair samples with different distribution of pair mass ratios yield different results, e.g. \citet{Hani-2020}. We emphasize that our SFR and metallicity measurements extend out to separations of $150\, \mathrm{kpc}$, whereas the sample of \citet{Scudder-2012} had a maximum projected separation of $80\, \mathrm{kpc}$, but found SFR enhancements still persisting at these wide separations.  Our sample is therefore better able to capture offsets at wide separations, see also \citet{Patton-2016}.

\subsubsection{Testing predictions of the FMR}\label{SecFMRTest}

In this subsection, we now extend our analysis by using the FMR described in equation~\ref{eqn:FMR} and the measured values of stellar mass and SFR for each galaxy pair, to compute the FMR-predicted metallicities (green-dashed line in the bottom panel of Figure~\ref{fig:DilutionPairs}). Similarly, we predict the SFR-enhancement based on pair's stellar mass and metallicity, also based on our characterisation of the FMR, in the top panel of Figure~\ref{fig:DilutionPairs}.

The FMR qualitatively predicts a dilution of metallicity for small projected separation ($r_p\lesssim$ 50 kpc), since the pairs also have larger positive SFR offsets with respect to the controls. Nevertheless, the measured metallicity offset is stronger than the FMR-prediction. At fixed stellar mass \emph{and} SFR, close galaxy pairs hence have lower metallicities than isolated galaxies and are therefore outliers on the FMR\footnote{The FMR is fitted to our general sample of galaxies, in which only about $10\%$ are potentially interacting galaxies. Therefore, the FMR encodes, on average, the behaviour of isolated galaxies.}. The FMR-predicted SFR enhancement for pairs at small separations (see upper panel of Figure~\ref{fig:DilutionPairs}) is also inconsistent with the measured SFR.

At intermediate projected separations $60\, \mathrm{kpc}<r_p<100\, \mathrm{kpc}$, a visual inspection shows that FMR-predicted metallicity offsets tend to be slightly larger in comparison to the measured offsets; however, the differences are small and statistically indistinguishable. In the same range of separations the SFR enhancement is significantly larger for the measured values compared to the FMR-predictions. Such a significant difference shows that the behaviour of galaxy pairs at intermediate distances cannot be captured by the FMR.

At large separations of $r_p>110$ kpc, the measured metallicity dilution and SFR enhancement agree with the FMR predictions. The behaviour of relatively wide pairs is therefore well captured by the FMR.

As mentioned above, Figure~\ref{fig:DilutionPairs} also shows the the median offsets of SFR and metallicity for post-merger galaxies. The values are $0.31\, \mathrm{dex}$ and $-0.064\, \mathrm{dex}$, respectively. These offsets are consistent with the slope of the relations for galaxy pairs at close projected separations, thereby suggesting that SFR enhancement and metallicity dilution still continue during post-merger stages. Although the FMR also predicts metallicity dilution in this case, i.e. a median of value of $-0.024\, \mathrm{dex}$, measured offsets are still on average lower than the measured values. After applying a Kolmogorov-Smirnov test to compare the distributions of observed metallicities vs FMR-predicted values, the hypothesis of both sets of values being drawn from the same distribution can be rejected at a confidence level of $\alpha = 2\%$. This implies that the population of post-merger galaxies is not well described by the FMR. B18 obtain similar results in simulated galaxies, in which SFR enhancement and metallicity dilution can last up to $300\, \mathrm{Myr}$ after coalescence and the FMR under-predicts the strength of the dilution.

All offsets, including those of FMR-predicted quantities, are calculated with respect to the statistical mean of measured values of control galaxies, i.e. by using equation~\ref{eqn:FractionalChange}. For the sake of consistency, we also compute the offsets of the FMR predictions for control galaxies. The corresponding medians are shown as red dotted lines in Figure~\ref{fig:DilutionPairs}. Deviations of metallicity offsets from the zero line are within a sensible range according to the fitted Gaussian distribution in Figure~\ref{fig:Residuals}. This shows that the FMR encodes the behaviour of isolated galaxies remarkably well.

\subsection{ Stellar mass ratio dependence } \label{SubsecMassRat}

Although both major and minor mergers have an important contribution to the galaxy assembly process, they are very different in nature. For example, whereas major mergers are much more disruptive, minor mergers are much more frequent \citep{Hani-2020}. Owing to this, it is interesting to study the difference in SFR and metallicity trends for major versus minor mergers. In order to do so, we define three different bins of merger mass ratio for the galaxy pair sample. A first bin of low mass companions in minor mergers ($\mu\leq 1/3$), with $3566$ galaxies ($47\, \%$ of the sample); a second bin of major mergers, both members included ($1/3<\mu \leq 3$), with $3374$ galaxies ($45\, \%$ of the sample); and a third bin of main galaxies in minor mergers ($\mu> 3$), with $574$ galaxies ($8\, \%$ of the sample). Unlike the parent sample of \citet{Patton-2016}, which has a mass ratio distribution that is fairly symmetric around $\mu = 1$, our sample is skewed towards relatively small mass ratios.
This is due to the fact that the higher mass galaxies in our sample are less likely to meet the emission line selection criteria outlined
in Section~\ref{SubsecGalSam}. The median masses of galaxies in the mass ratio bins are: $10^{9.6}\, \mathrm M_{\odot}$, $10^{9.9}\, \mathrm M_{\odot}$ and $10^{10.1}\, \mathrm M_{\odot}$, respectively. The same bins are also adopted by \citet{Scudder-2012}.

\begin{figure}
\centering
\includegraphics[width=0.35\textheight]{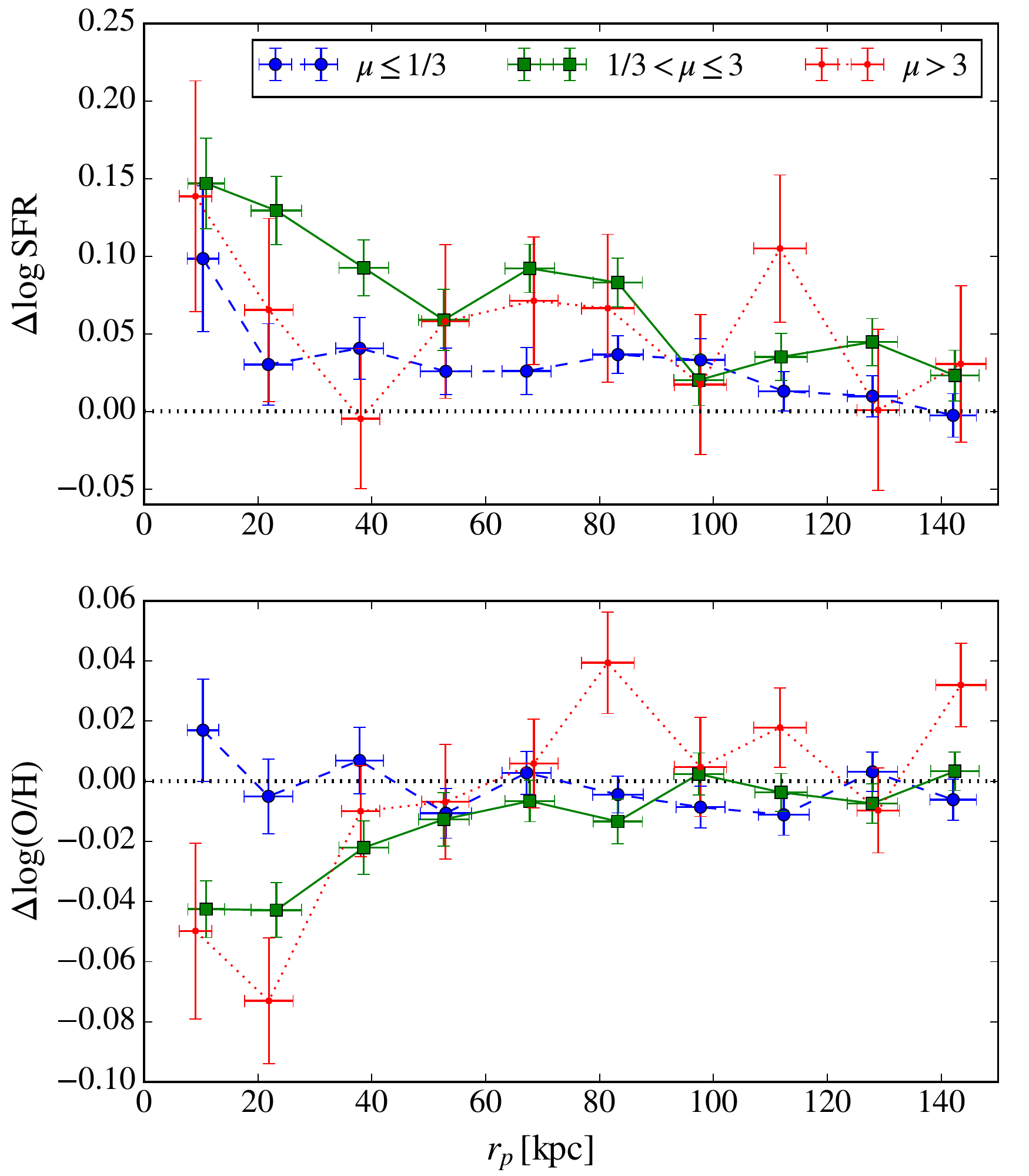}
  \caption{ Median offsets of SFR (\emph{top panel}) and metallicity (\emph{bottom panel}) as a function of projected separation for different bins of merger mass ratio. We define three different bins as follows: a first bin of low mass companions of minor mergers $1/10\leq \mu \leq 1/3$ (dashed lines and blue circles), a second bin of both members in major mergers $1/3<\mu\leq 3$ (solid lines and green squares), and third bin of main galaxies in minor mergers $3<\mu\leq 10$ (dotted lines and red dots). }
  \label{fig:DilutionPairs_StellarMassRatio}
\end{figure}

We plot the median offsets of SFR and metallicity with respect to the control sample as a function of projected separation in Figure~\ref{fig:DilutionPairs_StellarMassRatio}. We notice that SFR enhancement occurs in all cases, with major mergers expectedly exhibiting the higher offsets. Interestingly, the smaller companions in minor mergers experience significant SFR enhancements only at small separations $r_p\lesssim 20\, \mathrm{kpc}$, while main companions, with the exception of the bin at $40\, \mathrm{kpc}$, show relatively higher SFR enhancements in a wider range of separations. For metallicity offsets, we see that dilution only occurs appreciably in major mergers and massive companions in minor mergers, with the latter exhibiting slightly stronger dilutions at close separations. Smaller companions in minor mergers exhibit comparatively flatter profiles and lower offsets, implying thus that no significant dilution takes place. This result is consistent with \citet{2008MNRAS.386L..82M}, who observed a similar lack of metallicity dilution (they indeed observed an excess in metallicity) of the minor companion in a minor merger.

A correlation between galaxy mass and mass ratio is expected, given the finite range of stellar masses in our sample.  
For example, for galaxies in the low (high) stellar mass tail of our sample (see Fig.~\ref{fig:ControlHistograms}), most of the detectable companions will have higher (lower) mass, yielding smaller (larger) mass ratios. By randomly culling galaxies in the first bin of small companions in minor mergers and in the second bin of major mergers, we match the mass distribution of galaxies in the bin of massive companions in minor mergers. We use these sub-samples in order to recompute the median offsets as function of projected separation. Comparable trends to those in Figure~\ref{fig:DilutionPairs_StellarMassRatio} are obtained, i.e. major mergers and massive companions in minor mergers exhibit stronger offsets than small companions. This demonstrates that the strengths of the SFR enhancement and the metallicity dilution are less dependent on galaxy mass and indeed driven mostly by the mass ratio.

\subsection{ Environmental effects } \label{SubsecEnvEff}

\begin{figure}
\centering
\includegraphics[width=0.35\textheight]{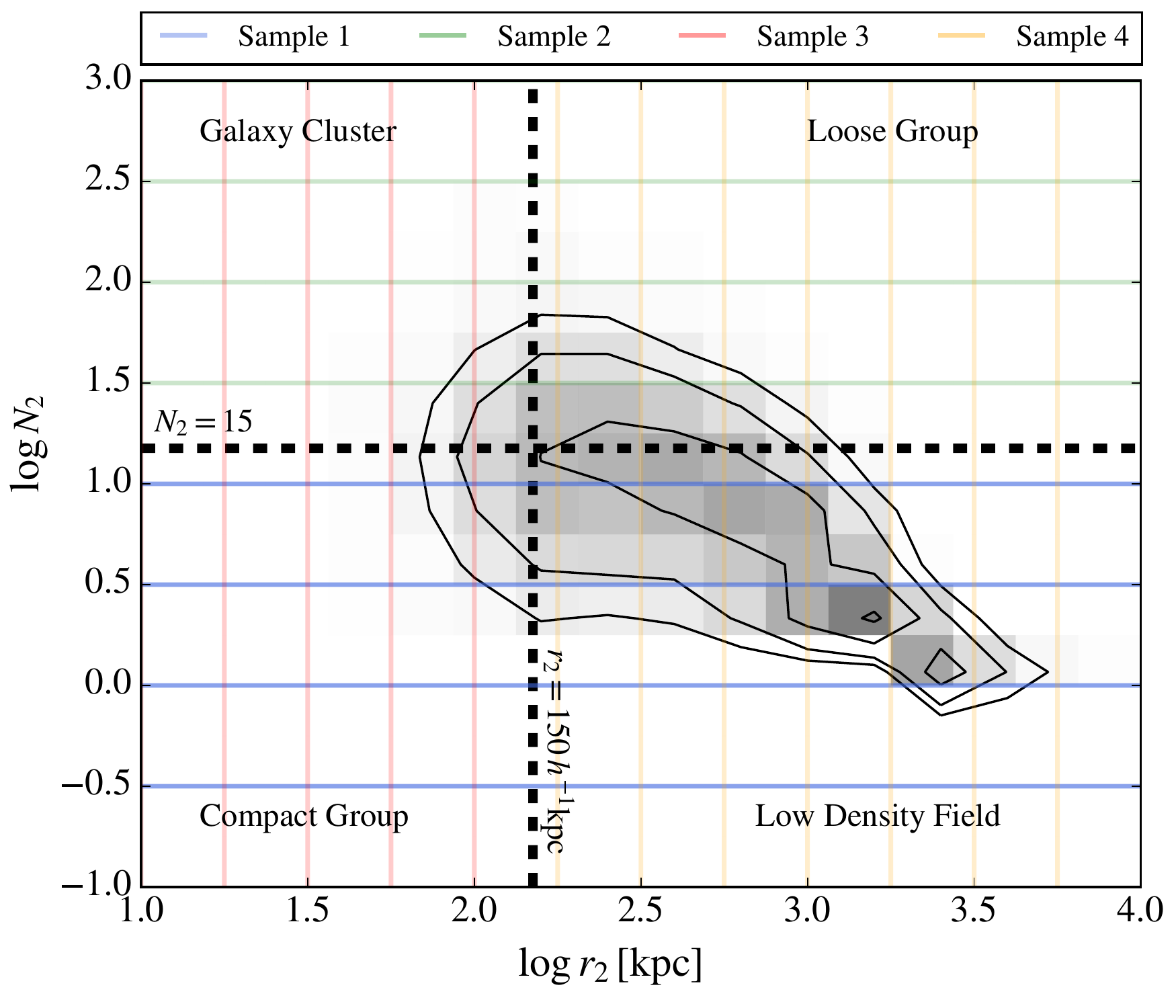}
  \caption{Environmental classification of galaxy pairs. Four different type of environments are defined in terms of $r_2$ and $N_2$, namely: compact groups, galaxy clusters, loose groups and low density field. The defining conditions are listed in subsection \ref{SubsecGalSam}. The 2D histogram represents the galaxy number count in each bin. Contour levels show galaxy number counts of $50$, $100$, $200$ and $400$. The majority of the galaxy pair sample is detected in low density fields. In order to quantify environmental effects on SFR enhancement and metallicity dilution, we define four samples represented by horizontal/vertical coloured lines.}
  \label{fig:EnvironmentClassification}
\end{figure}

Considering galaxy mergers as ideal isolated systems in hydrodynamical simulations has proven to be an accurate approximation as the effects exerted by close companions are, in most cases, more significant than cosmological effects; nevertheless, in dense environments and compact groups, other neighbouring galaxies might also play an important role in the evolution of galaxy properties \citep[B18]{Park-2007, Ellison-2010, Moreno-2013, Sabater-2013, Patton-2016, Sparre-2016}. In this subsection, we study the effects of the environment on the SFR enhancement and the metallicity dilution processes. In order to do so, we use the metrics $N_2$ and $r_2$ to define four different samples (see Figure~\ref{fig:EnvironmentClassification}). Sample~1 and sample~2 have different ranges of neighbour density, i.e. $N_2<15$ and $N_2\geq 15$, which include galaxies in compact groups and low density fields ($73\, \%$ of galaxy pairs), and galaxies in galaxy clusters and loose groups ($27\, \%$ of galaxy pairs), respectively. Sample~3 and sample~4 have different ranges of relative distance to a second neighbour, i.e. $r_2< 150\,   \mathrm{kpc}$ and $r_2\geq 150\,   \mathrm{kpc}$, which include galaxies in galaxy clusters and compact groups ($15\, \%$ of galaxy pairs), and galaxies in loose groups and low density fields ($85\, \%$ of galaxy pairs), respectively. 

\begin{figure}
\centering
\includegraphics[width=0.35\textheight]{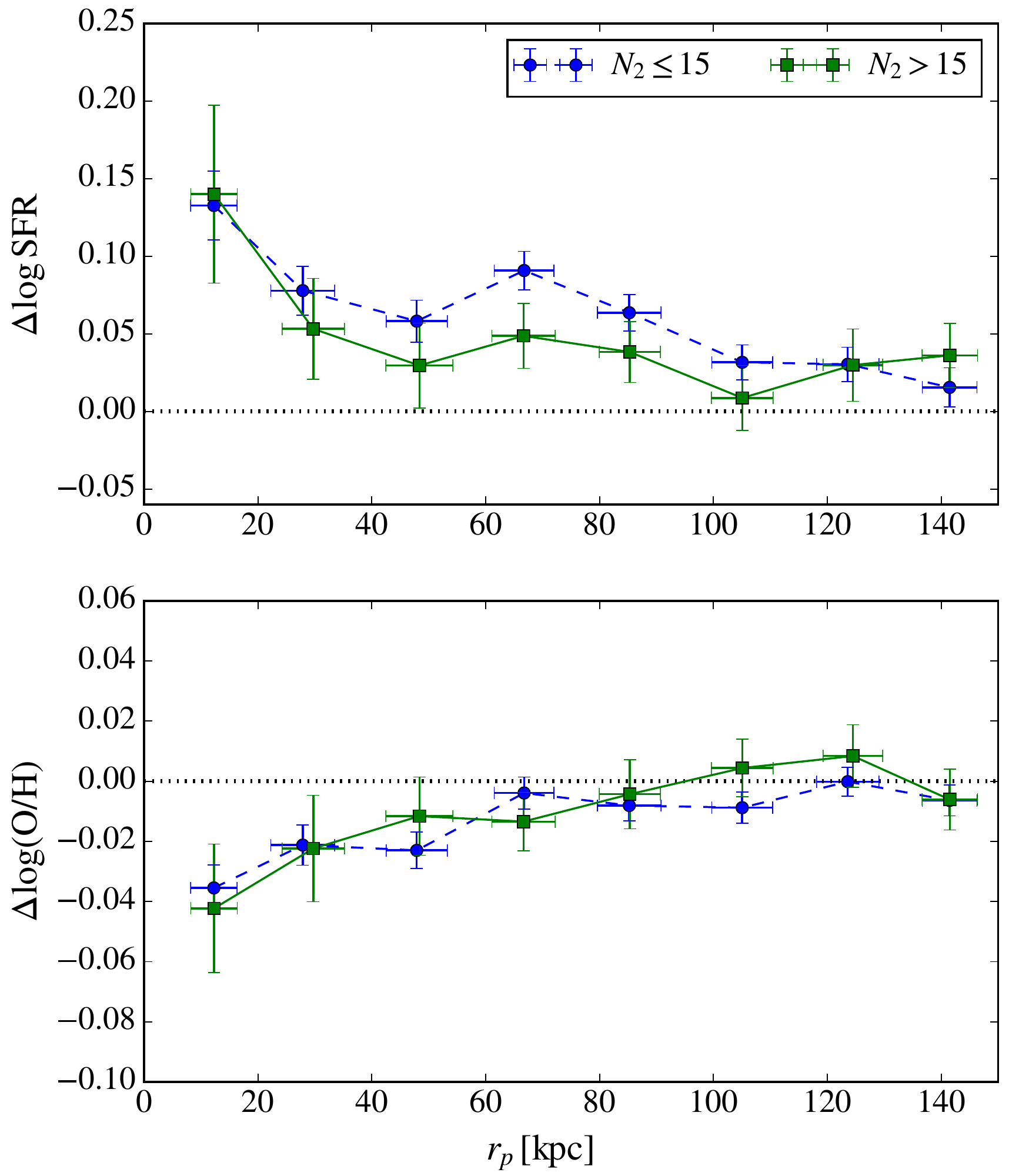}
  \caption{ Median offsets of SFR (\emph{top panel}) and metallicity (\emph{bottom panel}) as a function of projected separation for sample 1 ($N_2\leq 15$) and sample 2 ($N_2>15$). The use of these samples allows us to quantify the relative importance of the neighbour density $N_2$. We do not see an appreciable difference between the trends. }
  \label{fig:DilutionPairs_N2}
\end{figure}

In Figure~\ref{fig:DilutionPairs_N2}, we compute the median offsets of SFR and metallicity for galaxies in sample 1 and sample 2. By comparing the different trends, we are able to assess the relative importance of the neighbour density. We see no statistically significant differences in the median metallicity offsets. For SFR offsets, at projected separations $r_p > 40\,   \mathrm{kpc}$, galaxies in more dense environments have slightly lower values, i.e. providing evidence for environmental suppression of the SFR. 
We also compute the mass distributions of galaxies in these samples. The median masses of galaxies in sample 1 and sample 2 are $10^{9.78}\, \mathrm M_{\odot}$ and $10^{9.52}\, \mathrm M_{\odot}$, respectively. This difference indicates the presence of a selection effect in which less massive galaxies are preferentially detected in more dense environments. This is a consequence of the computation of $N_2$, in which only companions with stellar mass greater that $10\%$ of the host galaxy's stellar mass are considered. This means that for lower mass galaxies, $N_2$ probes deeper into the stellar mass function.  Since lower mass galaxies are more numerous, $N_2$ will typically be higher for these lower mass galaxies. In spite of this, the small statistical significance of the differences in metallicity and SFR offsets between these two samples -- particularly for metallicity -- shows that galaxy mass does not play an important role. This is in agreement with the results of subsection~\ref{SubsecMassRat} regarding the more relevant role of the mass ratio compared to the galaxy mass.

\begin{figure}
\centering
\includegraphics[width=0.35\textheight]{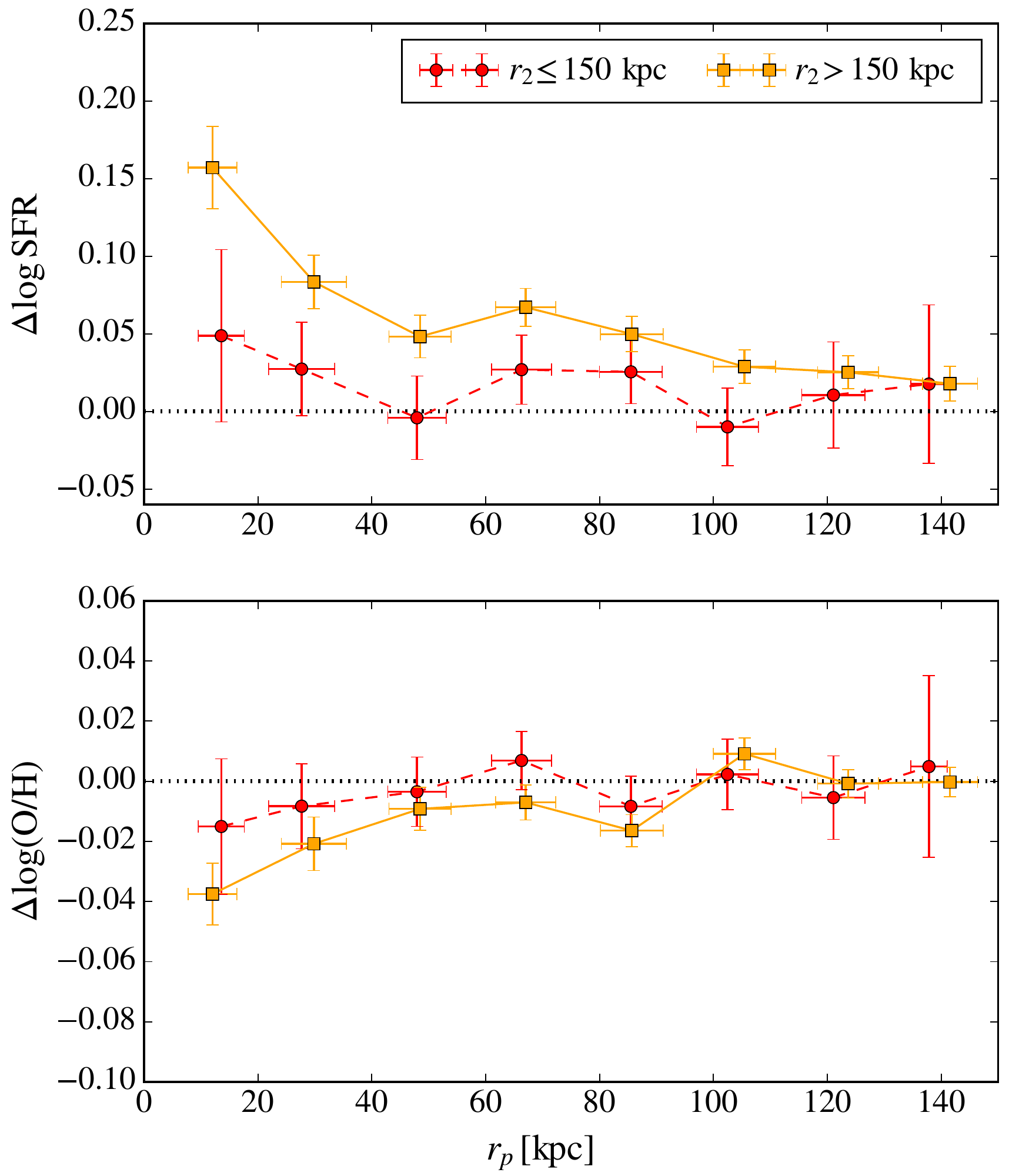}
  \caption{ Median offsets of SFR (\emph{top panel}) and metallicity (\emph{bottom panel}) as a function of projected separation for sample 3 ($r_2\leq 150  \mathrm{kpc}$) and sample 4 ($r_2> 150  \mathrm{kpc}$). The use of these samples allow us to quantify the relative importance of a second companion, i.e. through its projected separation $r_2$. SFR offsets of galaxies with a close second companion are, on average, lower than those of relatively isolated galaxies pairs. A weaker difference is also seen in metallicity offsets. }
  \label{fig:DilutionPairs_r2}
\end{figure}

We also compute median offsets for sample 3 and sample 4 in Figure \ref{fig:DilutionPairs_r2}. In this case, we want to assess the importance of a second companion in the SFR enhancement and the metallicity dilution in galaxy pairs. For the SFR, galaxy pairs with a close second companion exhibit smaller offsets than more isolated pairs. Likewise, metallicity offsets in isolated pairs are also slightly larger. The mass distributions of galaxies in sample 3 and sample 4 are very similar, with median values of $10^{9.62}\, \mathrm M_{\odot}$ and $10^{9.68}\, \mathrm M_{\odot}$, respectively. Selection effects between galaxy mass and projected separation of the second closest companion are therefore small or nonexistent in the galaxy pair sample, showing thus that the differences in the trends are indeed caused by the presence of a second companion.

For all the samples, the offsets have a strong dependence on the projected separation to the closest galaxy, thereby demonstrating that the interaction with a close companion is what drives the SFR enhancement and metallicity dilution processes. The environmental effects become relevant only through the presence of a close second companion. The local density of neighbouring galaxies does not seem to play an appreciable role.

\section{Discussion}\label{SecDiscussion}

The existence of an FMR, first discovered over a decade ago by \citet{Ellison-2008}, is supported by numerous observational works \citep[e.g.][]{Mannucci-2010, Lara-Lopez-2010, Belli-2013, Stott-2013,Salim-2014, Yabe-2015}.  The existence of an FMR is often understood in the context of gas supply and the subsequent impact of average chemical enrichment and the response of star formation.  Consequently, it has been suggested that the fundamental governing relationship for metallicity at fixed mass is based on gas content, rather than SFR \citep{Bothwell-2013,Bothwell-2016}.  Indeed, galaxies with elevated SFRs for their mass also have higher gas fractions \citep[e.g.][]{Saintonge-2016} and are more effective at converting that gas into stars \citep{Saintonge-2012,Piotrowska-2019}, a trend which seems to hold down to at least kpc-scales \citep{Ellison-2020}. 

It is observationally well established that interacting galaxies exhibit elevated SFRs and low gas phase metallicities, compared with stellar mass matched control samples \citep{Ellison-2008B, Rupke-2010, Scudder-2012, Cortijo-Ferrero-2017, Thorp-2019}. Moreover, galaxies in mergers exhibit elevated atomic and molecular gas fractions compared with isolated galaxies \citep{Ellison-2015-PMHI,Ellison-2018-PMHI, Violino-2018}.  The observed trends of metallicity, gas content and SFR, are therefore qualitatively consistent with the general galaxy population, as also expected from simulations of galaxy mergers \citep[e.g.][]{Torrey-2012, Hani-2018, Moreno-2019}. None the less, a different distribution may be expected when the processes of accretion and star formation happen more rapidly \citep{Mannucci-2010}, and an outlying population that is too metal-poor for its mass and SFR has been identified and proposed to be linked to mergers \citep{Gronnow-2015}.

In this paper, our goal has been to explicitly test, using a large sample of galaxies in both the pre- and post-coalescence phase, whether mergers contribute additional scatter to the FMR.  We have shown that the offsets of SFR and metallicity in SDSS galaxy pairs are not quantitatively consistent with FMR predictions.

Conceptual explanations for the existence of an FMR are often based on equilibrium models that involve changes in gas availability and the corresponding response in SFR \citep[e.g.][]{Dave-2012,Lilly-2013}. However, the time-scale on which these changes occur, and their coordination may play an important role in driving the strength and scatter of the relationship \citep{Forbes-2014,Torrey-2018}.  Therefore, whilst the same qualitative principles may be responsible for the hand-in-hand coordination of low O/H with high SFRs in mergers as in the general galaxy population, the prompt capture of these processes in on-going/recent interactions may explain the more extreme nature of this anti-correlation.  We conjecture that the impact of strong nuclear inflows (which may be short-lived) on diluting the gas phase metallicity is not captured by the generalised form of the FMR. As a result, the metallicities of galaxies currently undergoing an interaction, or those that have recently completed a merger, are offset to lower values of O/H than would nominally be predicted by the FMR. We conclude that interacting galaxies constitute a well-defined outlier population and that the FMR is not universal.

It is reassuring for the understanding of the FMR and the metallicity dilution in mergers that our observation-based result is qualitatively consistent with the analysis of cosmological merger simulations from B18, where the metallicity dilution during a merger was found also to be stronger than the FMR prediction. A possible direction for future analysis would be to track the low-metallicity gas during the coalescence of a cosmological merger simulation, to see how this is fuelled to the galaxy center from the circum-galactic medium or the outer part of the galaxy disc. If fuelling occurs on a shorter time-scale compared to the typical time-scale characterising the self-regulated processing giving rise to the FMR, we do indeed have a quantitative explanation for why the metallicity of major mergers are not captured by the FMR. In future work we plan to study this in new versions of the cosmological major merger simulations from \citet{Sparre-2017}, where we also include Monte Carlo tracer particles \citep[as described in][]{2013MNRAS.435.1426G}, which enable such an analysis.

In addition to the general offset of galaxy mergers from the FMR, we have investigated additional dependences on mass ratio and environment.  Major mergers and the more massive galaxy in a minor merging pair show significant offsets from the FMR, whereas lower mass companions show little deviation (Figure~\ref{fig:DilutionPairs_StellarMassRatio}).
We hypothesise that the small offsets of metallicity and SFR for the less massive galaxy companions in minor mergers are caused by the strangulation mechanism described by \citet{Peng2015}. In this scenario, the supply of cold gas in the galaxy is halted as a result of tidal stripping by the massive companion, thereby preventing a boost in star formation activity. Only at small separations comparable to the tidal radius of the small companion, the SFR is enhanced as gas accretion from the massive companion is possible. 

A similar mechanism could explain the smaller offsets of galaxy pairs with a close third galaxy compared to those with a distant one (see Figure~\ref{fig:DilutionPairs_r2}). The presence of a close third galaxy might result in tidal stripping of the cold gas supply of the pair. On the other hand, other environmental conditions such as the local density of neighbouring galaxies does not seem to play an appreciable role. We note that the extent of the effect of a third galaxy would depend on other parameters like the mass of the companion and the specific orbital configuration. Further analysis is needed in order to confirm this scenario.

\section{Summary and Conclusions}\label{SecSummary}

In this paper, we study the processes of metallicity dilution and SFR enhancement in interacting galaxies. Specifically, our goal is to assess whether the well-known metallicity dilution and SFR enhancement observed during the pair and post-merger phase is quantitatively consistent with the fundamental metallicity relation.

The galaxy merger sample is selected from the SDSS DR7 and consists of both spectroscopically selected galaxy pairs and visually identified post-mergers.  For each of the galaxies in these merger samples, we identify 10 control galaxies that are matched in stellar mass, redshift and local environment (as quantified by near neighbour statistics).  The FMR is defined using a large parent sample of galaxies from the SDSS DR7, and parametrized using a second order polynomial fit to stellar mass, SFR and metallicity. We use this second order characterisation of the FMR to predict O/H (given the stellar mass and SFR), or alternatively to predict SFR (given the stellar mass and O/H) for each galaxy in the merger and control samples.  The difference between the predicted O/H (or SFR) and measured value allows us to quantify how well the FMR characterises the merger population (and its controls).

Our main result is that galaxy mergers, identified in both the pre- and post-coalescence phase, are quantitatively offset from the FMR.  The offset is correlated with pair phase, increasing as the galaxies get closer together and peaking in the post-merger regime. Quantified in terms of star formation excess, post-mergers have SFRs that are $\sim$ 0.65 dex higher than predicted by the FMR, for their masses and metallicities. Quantified in terms of metallicity, post-mergers are $\sim$ 0.04 dex too metal-poor to lie on the FMR. Control galaxies, on the other hand, are statistically consistent with the FMR. Furthermore, we conclude that only galaxy pairs with separation larger than 110 kpc have a metallicity and SFR quantitatively in agreement with the FMR.

We suggest that the principles often used to explain the existence of the FMR, namely the regulation between gas availability and star formation, are still at play in galaxy mergers.  Indeed, many simulations have demonstrated the connection between gas inflows and central star formation. However, compared to the general galaxy population, whose changes in gas and SFR may be slow and hence rather coordinated, galaxy mergers represent a regime in which gas inflows are both recently triggered and are particularly strong. As a result, the anti-correlation between SFR and O/H may be more extreme than in the local galaxy population.

\section*{Acknowledgements}

SB acknowledges support from the International Max-Planck Research School for Astronomy and Cosmic Physics of Heidelberg (IMPRS-HD) and financial support from the Deutscher Akademischer Austauschdienst (DAAD) through the program Research Grants - Doctoral Programmes in Germany (57129429) and the Klaus Tschira Foundation. SLE and DRP acknowledge receipt
of Discovery Grants from NSERC of Canada. MS acknowledges support by the European Research Council under ERC-CoG grant CRAGSMAN-646955. We have used NASAs ADS Bibliographic Services. Most of the computational analysis was performed with {\tt Python 2.7} and its related tools and libraries, {\tt iPython} \citep{Perez-2007}, {\tt Matplotlib} \citep{Hunter-2007}, {\tt scipy} and {\tt numpy} \citep{Van-2011}. 

We thank the referee for many constructive suggestions that improved the clarity and presentation of this paper.

The SDSS is managed by the Astrophysical Research Consortium for the Participating Institutions. The Participating Institutions are the American Museum of Natural History, Astrophysical Institute Potsdam, University of Basel, University of Cambridge, Case Western Reserve University, University of Chicago, Drexel University, Fermilab, the Institute for Advanced Study, the Japan Participation Group, Johns Hopkins University, the Joint Institute for Nuclear Astrophysics, the Kavli Institute for Particle Astro-physics and Cosmology, the Korean Scientist Group, the Chinese Academy of Sciences (LAMOST), Los Alamos National Laboratory, the Max-Planck-Institute for Astronomy (MPIA), the Max-Planck-Institute for Astrophysics (MPA), New Mexico State University, Ohio State University, University of Pittsburgh, University of Portsmouth, Princeton University, the United States Naval Observatory and the University of Washington.

\footnotesize{
\bibliographystyle{mnras}
\bibliography{paper}
}

\bsp	
\label{lastpage}
\end{document}